\documentclass[11pt]{article}
\usepackage[a4paper,margin=1in]{geometry}
\usepackage[T1]{fontenc}
\usepackage{lmodern}
\usepackage{microtype}
\usepackage{amsmath,amssymb,mathtools}
\usepackage{bm}
\usepackage{siunitx}
\usepackage{physics}
\usepackage{graphicx}
\usepackage{xcolor}
\usepackage[hidelinks]{hyperref}
\usepackage[capitalize,nameinlink]{cleveref}
\usepackage[numbers,sort&compress]{natbib}
\usepackage{amsthm}

\theoremstyle{plain}
\newtheorem{proposition}{Proposition}
\AtBeginDocument{\RenewCommandCopy\qty\SI} 
\newcommand{\trop}{\mathrm{trop}}
\newcommand{\hz}{\hat{\mathbf{z}}}
\newcommand{\bfx}{\mathbf{x}}
\newcommand{\bfU}{\mathbf{U}}
\newcommand{\D}{\mathrm{D}}

\newcommand{\gradH}{\nabla_{\!h}}

\newcommand{\p}{\partial}

\newcommand{\J}{\mathcal{J}}

\title{Stratosphere Model Verification with Manufactured Geometry}
\author{Johannes~\textsc{Lawen} \and George~\textsc{Salman} \and Akshita~\textsc{Bhardwaj}}
\date{Draft -- \today}

\begin{document}
\maketitle

\begin{abstract}
We propose a stratosphere-only hydrostatic dynamical core formulated in geopotential/pressure coordinates with a \emph{time-evolving lower boundary} supplied by the troposphere. Rather than constraining the stratospheric circulation via specified dynamics (``nudging'') to a reanalysis, we treat the tropopause as a \emph{moving geometric boundary}. The stratospheric domain thus expands, contracts, and undulates in response to tropospheric variability while preserving familiar hybrid $\sigma$--$p$ structure and pressure-gradient calculations. The approach integrates naturally with arbitrary Lagrangian--Eulerian (ALE) updates and conservative remap to maintain positive layer thickness and tracer monotonicity. We outline the formulation, highlight analytical properties (well-posedness, energetics, wave propagation), and sketch a verification/validation path based on modified standard test cases and reanalysis-driven experiments.
\end{abstract}

\section{Introduction}
\label{sec:intro}
The dynamics and composition of the stratosphere modulate surface climate, ozone recovery, and extremes through a web of teleconnections and wave--mean flow interactions\citep{Brewer1949,Dobson1956,AndrewsHoltonLeovy1987,CharneyDrazin1961}. A common way to study such processes in isolation is to run ``stratosphere-only'' configurations in which the large-scale circulation is constrained---or \emph{specified}---by relaxing toward a reanalysis product\citep[e.g.,][]{Hersbach2020,Gelaro2017}. This practice is effective and widely adopted, yet it inevitably carries the imprint of data assimilation and mixes tropospheric and stratospheric variability at the level of the \emph{prognostic flow}, complicating attribution and mechanistic interpretation.

We take a different route that keeps the stratospheric equations of motion prognostic while coupling to the troposphere kinematically through \emph{geometry}. In geopotential/pressure (or hybrid $\sigma$--$p$) coordinates\citep{SimmonsBurridge1981,Lin2004}, the location of the lower boundary can vary in space and time without disturbing hydrostatic balance or the structure of the horizontal pressure-gradient force. If we identify this moving boundary with the tropopause, $z_{\trop}(x,y,t)$, or, equivalently, with a time-dependent $p_{\trop}$ surface, the troposphere becomes a \emph{dynamic bottom} that deforms the stratospheric domain---akin to evolving bathymetry/topography in geophysical fluids. The stratospheric core remains compact and hydrostatic; tropospheric influence enters purely through boundary motion.

To make this concrete, let $\eta\in[0,1]$ be a generalized vertical coordinate with the hybrid mapping
\begin{equation}
  p(\eta,\bfx,t) = A(\eta)\,p_0 + B(\eta)\,p_{\trop}(\bfx,t),
  \label{eq:hybrid}
\end{equation}
where $A(0)=0$, $B(0)=0$, and $B(1)=1$, so that $\eta=1$ tracks the moving lower boundary while the top is fixed at $p(0)=p_\text{top}$. Hydrostatic balance, $\p\Phi/\p p = -\alpha$, is preserved, as are standard geopotential-based pressure-gradient formulations\citep{SimmonsBurridge1981}. The kinematic condition at the boundary is
\begin{equation}
  w\big|_{\eta=1} = \frac{\D z_{\trop}}{\D t} = \p_t z_{\trop} + \bfU\cdot\gradH z_{\trop},
  \label{eq:bc}
\end{equation}
with $\bfU$ the horizontal wind evaluated at $\eta=1$. Vertical-level motion and mesh quality are handled by an arbitrary Lagrangian--Eulerian (ALE) update followed by a conservative, monotone remap, adopting mature techniques from ocean and compressible-flow models\citep{HirtAmsdenCook1974,Bleck2002,LinRood1996}. When $z_{\trop}$ is held fixed, the formulation reduces exactly to a standard hybrid $\sigma$--$p$ core, providing a strong verification hook.

Treating the tropopause as a dynamic surface cleanly separates the stratospheric circulation from reanalysis assimilation artifacts while still passing tropospheric influence through domain geometry. This perspective naturally preserves balances and energetics: work done by boundary motion enters the budget through well-defined geometric flux terms that can be diagnosed alongside wave activity and mean-flow acceleration. Conceptually, it reframes lower-stratospheric sensitivity to tropospheric variability as a boundary-value problem, opening the door to normal-mode and WKB analyses of wave propagation across an undulating interface.

Positioned within the literature, our approach links three strands: (i) terrain-following and hybrid vertical coordinates for primitive-equation models\citep{SimmonsBurridge1981,Lin2004}; (ii) specified-dynamics studies that drive the stratosphere by relaxing the flow toward reanalysis\citep{Hersbach2020,Gelaro2017}; and (iii) ALE/moving-boundary numerics that maintain positive layer thickness and tracer monotonicity in evolving meshes\citep{HirtAmsdenCook1974,Bleck2002,LinRood1996}. To our knowledge, prior stratosphere-only work transmits tropospheric influence by nudging the \emph{circulation}; here we transmit it kinematically via a \emph{moving domain}. This provides a complementary route to interrogate wave--mean flow interaction and the Brewer--Dobson circulation\citep{Brewer1949,Dobson1956,AndrewsHoltonLeovy1987} without relaxing the dynamical core itself.
Our contributions are threefold. First, we present a hydrostatic primitive-equation stratospheric model in geopotential/pressure coordinates with a time-varying lower boundary $z_{\trop}(x,y,t)$ that collapses to the standard hybrid $\sigma$--$p$ core when $\p_t z_{\trop}\equiv 0$. 
Second, we develop a perturbative framework for small-amplitude boundary undulations---including normal modes and WKB scalings---and an energetics budget in which boundary work appears explicitly, linking geometric forcing to changes in wave activity and mean-flow acceleration. Third, we outline a verification/validation path that adapts canonical tests (e.g., the Jablonowski--Williamson baroclinic wave\citep{JablonowskiWilliamson2006}) to a moving boundary and complements them with reanalysis-driven experiments in which $z_{\trop}(x,y,t)$ is prescribed from ERA5/MERRA-2 while the stratospheric flow remains prognostic.

We proceed as follows. We first define the equations, coordinate transform, and ALE/remap update. We then develop the perturbation theory and energy budget, followed by a discussion of discretization and stability. Finally, we present a verification suite and close with limitations and opportunities.
\subsection{Solution}
The hybrid pressure coordinate is defined by
\begin{equation}
  p(\eta,\mathbf{x},t)
  = A(\eta)\,p_0 + B(\eta)\,p_{\trop}(\mathbf{x},t),
  \qquad \eta\in[0,1],
\end{equation}
with $A(0)=0$, $B(0)=0$, and $B(1)=1$, so that the upper boundary is fixed and
$p(\eta=1)=p_{\trop}(\mathbf{x},t)$. Differentiation with respect to $\eta$
yields
\begin{equation}
  \frac{\partial p}{\partial \eta}
  = A'(\eta)\,p_0 + B'(\eta)\,p_{\trop}(\mathbf{x},t),
\end{equation}
which is strictly positive for admissible choices of $A$ and $B$, ensuring an
invertible coordinate transformation. Taking the material derivative of $p$
at fixed $\eta$ gives
\begin{equation}
  \frac{D p}{D t}\Big|_{\eta}
  = B(\eta)\,\frac{D p_{\trop}}{D t},
\end{equation}
since $A(\eta)$ and $B(\eta)$ are time independent. Hydrostatic balance in
pressure coordinates,
\begin{equation}
  \frac{\partial \Phi}{\partial p} = -\alpha,
\end{equation}
is preserved exactly under the transformation, since application of the chain
rule yields
\begin{equation}
  \frac{\partial \Phi}{\partial \eta}
  = \frac{\partial \Phi}{\partial p}
    \frac{\partial p}{\partial \eta}
  = -\alpha\left(A'(\eta)p_0 + B'(\eta)p_{\trop}\right).
\end{equation}
The vertical velocity is defined by the material derivative of height,
\begin{equation}
  w = \frac{D z}{D t}
    = \partial_t z + \mathbf{U}\cdot\nabla_H z
      + \dot{\eta}\,\partial_\eta z,
\end{equation}
which may be solved exactly for the coordinate velocity,
\begin{equation}
  \dot{\eta}
  = \frac{w - \partial_t z - \mathbf{U}\cdot\nabla_H z}
         {\partial_\eta z}.
\end{equation}
At the lower boundary $\eta=1$, where $z(\eta=1,\mathbf{x},t)=z_{\trop}(\mathbf{x},t)$,
the vertical velocity satisfies
\begin{equation}
  w\big|_{\eta=1}
  = \frac{D z_{\trop}}{D t}
  = \partial_t z_{\trop}
    + \mathbf{U}\cdot\nabla_H z_{\trop},
\end{equation}
so that the kinematic boundary condition is satisfied identically. If the
tropopause is fixed such that $\partial_t p_{\trop}=0$ and
$\nabla_H p_{\trop}=0$, then $D p/D t|_{\eta}=0$ and $\dot{\eta}=0$, and the
formulation reduces exactly to the standard hybrid $\sigma$--$p$ system.

\subsection{Incompressible case: 2D specialization (manufactured geometry)}
\label{sec:incomp-2d}

We consider a free-surface, incompressible shallow-flow analog in two horizontal
dimensions $(x,y)$ with eddy viscosity $k_H$ and gravitational acceleration $g$.
Let $h_E(x,y,t)$ denote surface elevation and $h_B(x,y,t)$ the bathymetric
depth so that the total column height is $h=h_E+h_B$. The standard conservative
form is
\begin{align}
\p_t h + \p_x(h u) + \p_y(h v) &= 0, \label{eq:sw-cont-2d}\\
\p_t(h u) + \p_x(h u^2) + \p_y(h u v) &= -\,g\,h\,\p_x h_E
+ k_H\!\left[\p_x\!\big(h\,\p_x u\big)+\p_y\!\big(h\,\p_y u\big)\right],
\label{eq:sw-momx-2d}\\
\p_t(h v) + \p_x(h u v) + \p_y(h v^2) &= -\,g\,h\,\p_y h_E
+ k_H\!\left[\p_x\!\big(h\,\p_x v\big)+\p_y\!\big(h\,\p_y v\big)\right].
\label{eq:sw-momy-2d}
\end{align}

\paragraph{Manufactured fields.}
Let $\Theta(x,y,t)\equiv x+y+t$ and choose a constant $c_1>1$ to keep $h$ positive.
Define
\begin{equation}
h(x,y,t) = c_1 + \sin\Theta,\qquad
u(x,y,t) = \frac{1}{\sin\Theta+c_1}\;-\;\frac{1}{2},\qquad
v(x,y,t) = \frac{1}{\sin\Theta+c_1}\;-\;\frac{1}{2}.
\label{eq:ansatz-2d}
\end{equation}
With \eqref{eq:ansatz-2d}, the continuity equation \eqref{eq:sw-cont-2d} holds identically.

\paragraph{Partition into $h_E$ and $h_B$.}
Write $h=h_E+h_B$ and treat $h_E$ as the unknown “free surface’’ to be balanced
by a manufactured $h_B$. Substituting \eqref{eq:ansatz-2d} into
\eqref{eq:sw-momx-2d}–\eqref{eq:sw-momy-2d} gives the required horizontal
pressure gradients as functions of $\Theta$,
\begin{align}
\p_x h_E &= \mathcal{R}(\Theta),\qquad
\p_y h_E = \mathcal{R}(\Theta), \label{eq:he-grad}\\
\mathcal{R}(\Theta) &=
-\frac{1}{g}\Bigg[
\p_t(hu) + \p_x(hu^2) + \p_y(huv)
- k_H\!\left(\p_x\!\big(h\,\p_x u\big)+\p_y\!\big(h\,\p_y u\big)\right)
\Bigg]\!, \nonumber
\end{align}
where each derivative reduces to an ordinary derivative in $\Theta$ since all
fields depend on $(x,y,t)$ only through $\Theta$. Because
$\p_x\Theta=\p_y\Theta=1$, the compatibility condition
$\p_y(\p_x h_E)=\p_x(\p_y h_E)$ is automatically satisfied and there exists a
scalar potential $H(\Theta)$ with $h_E(x,y,t)=H(\Theta)$ and $H'(\Theta)=\mathcal{R}(\Theta)$.
A convenient antiderivative is
\begin{equation}
H(\Theta) = \int^{\Theta}\!\mathcal{R}(s)\,\mathrm{d}s
\quad\Rightarrow\quad
h_E(x,y,t) = H\!\big(\Theta(x,y,t)\big),
\qquad
h_B(x,y,t) = h - h_E. \label{eq:hE-hB-sol}
\end{equation}

\paragraph{Closed form for $\mathcal{R}(\Theta)$.}
With \eqref{eq:ansatz-2d}, denote $S(\Theta)\equiv\sin\Theta$ and
$C(\Theta)\equiv\cos\Theta$ for brevity. Then
\begin{align*}
h &= c_1+S, & u=v &= \frac{1}{c_1+S}-\frac{1}{2}, &
\p_\Theta h &= C, &
\p_\Theta u=\p_\Theta v &= -\frac{C}{(c_1+S)^2}.
\end{align*}
A straightforward calculation yields
\begin{align}
\mathcal{R}(\Theta)
&= -\frac{1}{g}\Bigg\{
C\left(\frac{1}{c_1+S}-\frac{1}{2}\right)
+\p_\Theta\!\left[(c_1+S)\!\left(\frac{1}{c_1+S}-\frac{1}{2}\right)^{\!2}\right]
\nonumber\\
&\hspace{6em}
-\,k_H\,\p_\Theta\!\left[(c_1+S)\,\frac{C}{(c_1+S)^2}\right]
\Bigg\}. \label{eq:R-theta}
\end{align}
Integrating \eqref{eq:R-theta} in \eqref{eq:hE-hB-sol} gives an explicit $H(\Theta)$
(up to an arbitrary additive constant). In particular, the elementary parts
produce terms proportional to $\sin\Theta$, $\ln|c_1+\sin\Theta|$, and $(c_1+\sin\Theta)^{-1}$.
For completeness,
\begin{equation}
h_B(x,y,t)=c_1+\sin\Theta-H(\Theta),\qquad
\Theta=x+y+t.
\end{equation}

\paragraph{Remarks.}
(i) Choosing $c_1>1$ avoids zeros in $c_1+\sin\Theta$ and maintains $h>0$.
(ii) The construction is scalar and portable: replacing $\Theta$ by
$k_x x + k_y y + \omega t$ simply scales the derivatives by constants.
(iii) This manufactured triplet $(h,u,v)$ together with $h_B$ from
\eqref{eq:hE-hB-sol} satisfies \eqref{eq:sw-cont-2d}–\eqref{eq:sw-momy-2d}
by construction and is suitable for FV/FE regression tests on arbitrary meshes.

\subsection{Compressible Case}
\[
\textbf{Conserved quantities:} \quad 
\mathbf{U} =
\begin{pmatrix}
\rho h \\[4pt]
\rho h u \\[4pt]
\rho h v \\[4pt]
\rho h T
\end{pmatrix}
\]

\[
\textbf{Governing equations for two spatial dimensions in conservative form:}
\]
\[
\begin{aligned}
\frac{\partial(\rho h)}{\partial t}
&+ \frac{\partial(\rho h\,u)}{\partial x}
+ \frac{\partial(\rho h\,v)}{\partial y}
= 0,
\\[8pt]
\frac{\partial(\rho h\,u)}{\partial t}
&+ \frac{\partial(\rho h\,u^2)}{\partial x}
+ \frac{\partial(\rho h\,u v)}{\partial y}
= -\,h\,\frac{\partial p}{\partial x} + \rho h\,F_x,
\\[8pt]
\frac{\partial(\rho h\,v)}{\partial t}
&+ \frac{\partial(\rho h\,u v)}{\partial x}
+ \frac{\partial(\rho h\,v^2)}{\partial y}
= -\,h\,\frac{\partial p}{\partial y} + \rho h\,F_y,
\\[8pt]
\frac{\partial(\rho h\,T)}{\partial t}
&+ \frac{\partial(\rho h\,u T)}{\partial x}
+ \frac{\partial(\rho h\,v T)}{\partial y}
= \rho h\,Q.
\end{aligned}
\]
\[
\textbf{Flux vectors:}
\]
\[
\mathbf{F}_x =
\begin{pmatrix}
\rho h u \\[4pt]
\rho h u^2 \\[4pt]
\rho h u v \\[4pt]
\rho h u T
\end{pmatrix},
\qquad
\mathbf{F}_y =
\begin{pmatrix}
\rho h v \\[4pt]
\rho h u v \\[4pt]
\rho h v^2 \\[4pt]
\rho h v T
\end{pmatrix}.
\]

\[
\textbf{Source vector:}
\qquad
\mathbf{S} =
\begin{pmatrix}
0 \\[4pt]
-\,h\,\dfrac{\partial p}{\partial x} + \rho h F_x \\[6pt]
-\,h\,\dfrac{\partial p}{\partial y} + \rho h F_y \\[6pt]
\rho h Q
\end{pmatrix}.
\]

\[
\textbf{Compact vector form:} \qquad
\frac{\partial \mathbf{U}}{\partial t}
+ \frac{\partial \mathbf{F}_x}{\partial x}
+ \frac{\partial \mathbf{F}_y}{\partial y}
= \mathbf{S}.
\]

\[
\textbf{Equation of state (ideal gas):} \qquad
p = \rho R T,
\]
so that
\[
\frac{\partial p}{\partial x}
= R\!\left(T\,\frac{\partial \rho}{\partial x}
+ \rho\,\frac{\partial T}{\partial x}\right),
\qquad
\frac{\partial p}{\partial y}
= R\!\left(T\,\frac{\partial \rho}{\partial y}
+ \rho\,\frac{\partial T}{\partial y}\right).
\]

\section{Formulation overview (sketch)}
\label{sec:formulation}

We summarize the stratosphere-only dynamical core with a time-evolving lower
boundary. The goal is to (i) keep the interior primitive equations fully
prognostic in the stratosphere, (ii) represent tropospheric influence purely
through boundary motion at the tropopause, and (iii) do so in a way that
remains hydrostatic, conservative, and numerically stable.

\subsection{Coordinates, mapping, and prognostic state}
\label{sec:coords-mapping}

We employ the hybrid geopotential/pressure mapping already introduced in
\cref{eq:hybrid},
\begin{equation}
  p(\eta,\bfx,t)
  = A(\eta)\,p_0 + B(\eta)\,p_{\trop}(\bfx,t),
  \qquad \eta \in [0,1],
  \label{eq:hybrid-repeat}
\end{equation}
with $A(0)=0$, $B(0)=0$, $B(1)=1$, and $p(\eta{=}0)=p_{\text{top}}$ fixed.
By construction, $\eta=1$ coincides with the tropopause and therefore moves
in space and time with $p_{\trop}(\bfx,t)$ and $z_{\trop}(\bfx,t)$, while
$\eta=0$ is a fixed (high) model top.

In this coordinate, each horizontal column is discretized into $N_\eta$
layers bounded by $\eta_k$ surfaces. Because $p_{\trop}(\bfx,t)$ appears
explicitly in \cref{eq:hybrid-repeat}, the \emph{physical} pressure thickness
and geometric thickness of those layers change when the tropopause moves.

The prognostic state advanced on these $\eta$-levels consists of
\begin{itemize}
  \item the horizontal wind components $(u,v)$,
  \item temperature (or potential temperature),
  \item any tracers (e.g.\ ozone, water vapour),
  \item and the tropopause pressure field $p_{\trop}(\bfx,t)$.
\end{itemize}
Vertical momentum is diagnosed hydrostatically from
$\partial \Phi / \partial p = -\alpha$, with
$\alpha = R T / p$, as in a standard hydrostatic primitive-equation core
\citep{SimmonsBurridge1981,Lin2004}. No explicit prognostic vertical velocity
equation is needed; instead $w$ is obtained from mass continuity and the
kinematic lower-boundary condition \cref{eq:bc}.

It is convenient to denote the Jacobian of the $(\eta,\bfx)\mapsto(p,\bfx)$
mapping by
\begin{equation}
  \J(\eta,\bfx,t)
  \equiv \frac{\partial(p,\bfx)}{\partial(\eta,\bfx)}
  = \frac{\partial p}{\partial \eta}(\eta,\bfx,t),
  \label{eq:jacobian}
\end{equation}
since the horizontal coordinates are unchanged.
The layer ``mass per unit area'' is then proportional to $\J\,\Delta \eta$,
and changes in $\J$ capture the vertical mesh motion forced by the evolving
tropopause.

\subsection{Mass continuity in the moving vertical coordinate}
\label{sec:mass-cont}

Let $\bfU=(u,v)$ be the horizontal velocity and $w$ the physical vertical
velocity in geometric height $z$. In generalized coordinates tied to
\cref{eq:hybrid-repeat}, mass continuity can be written in flux form as
\begin{equation}
  \partial_t (\J)
  + \nabla_{\!h}\!\cdot(\J\,\bfU)
  + \partial_\eta (\J\,\dot{\eta})
  = 0,
  \label{eq:mass-cont-J}
\end{equation}
where $\dot{\eta} \equiv \mathrm{D}\eta/\mathrm{D}t$ is the ``vertical''
transport velocity in $\eta$-space, and $\nabla_{\!h}$ is the horizontal
gradient/divergence operator.

Equation~\eqref{eq:mass-cont-J} is just column mass conservation expressed
on a moving mesh: the first term is local tendency of layer mass, the second
is horizontal divergence of mass flux, and the third is the flux through
layer interfaces in $\eta$-space.

At $\eta=1$, the lower boundary is \emph{not fixed}: it is identified with
the tropopause. The exact kinematic boundary condition \cref{eq:bc} implies
that the mesh-following velocity at $\eta=1$ satisfies
\begin{equation}
  \dot{\eta}(\eta{=}1)
  \quad\Longleftrightarrow\quad
  w\big|_{\eta=1}
  = \frac{\mathrm{D}z_{\trop}}{\mathrm{D}t}
  = \partial_t z_{\trop}
    + \bfU\cdot\nabla_{\!h} z_{\trop}.
  \label{eq:bc-repeat}
\end{equation}
In other words, the ``vertical'' mass flux at the bottom of the stratospheric
column is entirely prescribed by the imposed geometry $z_{\trop}(\bfx,t)$.
This is the only way tropospheric variability enters the stratospheric mass
budget.

When $z_{\trop}$ (equivalently $p_{\trop}$) is \emph{held fixed}, then
$\partial_t z_{\trop}=0$ and \cref{eq:bc-repeat} reduces to the usual
impermeability condition $w|_{\eta=1}=0$. In that limit, the model reduces
exactly to a standard hybrid $\sigma$--$p$ core. This “fixed-bottom”
consistency provides an immediate verification path.

\subsection{ALE + conservative remap time step}
\label{sec:ale-remap}

We advance the model in three substeps following arbitrary
Lagrangian--Eulerian (ALE) ideas \citep{HirtAmsdenCook1974,Bleck2002} combined
with conservative remap \citep{LinRood1996,Lin2004}:

\begin{enumerate}
  \item \textbf{Mesh motion (ALE step).}
  Given $p_{\trop}^n(\bfx)$ (or $z_{\trop}^n$) at time $t^n$, we prescribe
  $p_{\trop}^{n+1}(\bfx)$ at $t^{n+1}$ from the chosen forcing
  (e.g.\ reanalysis, idealized oscillation). This defines a new tropopause
  surface and therefore a new mapping $p(\eta,\bfx,t^{n+1})$ via
  \cref{eq:hybrid-repeat}. Intermediate $\eta$-levels are displaced
  vertically as well, producing a ``moved'' mesh consistent with the updated
  tropopause.

  During this step, each layer is treated as moving vertically with a mesh
  velocity $w_{\text{mesh}}$ chosen such that $\eta$-layers do not cross and
  total column mass in each horizontal column is preserved. This is a
  purely vertical (columnwise) transport: there is no horizontal advection
  yet.

  \item \textbf{Conservative remap.}
  After mesh motion, the prognostic variables now live on the deformed,
  time-dependent $\eta$ grid. We remap them back onto a target grid
  (usually a slightly adjusted ``nice'' $\eta$ grid for $t^{n+1}$) using a
  conservative, monotone finite-volume remap in the spirit of
  \citet{LinRood1996}. This remap:
  (i) preserves cell-integrated mass,
  tracer mass, and (optionally) total energy to machine precision;
  (ii) applies slope limiting / PPM-style monotonicity constraints to avoid
  generating new extrema or negative layer thickness; and
  (iii) guarantees that no layer collapses to zero thickness even if
  $z_{\trop}$ has moved sharply.

  Conceptually, step (1) says “follow the tropopause so the lower boundary
  is represented exactly,” and step (2) says “re-project back to a regular
  mesh so numerics stay well behaved.”

  \item \textbf{Dynamical update.}
  Finally, on the remapped mesh at $t^{n+1}$ we solve the horizontal momentum,
  thermodynamic, and tracer equations (advection, Coriolis, pressure-gradient,
  and diffusion terms) with standard hydrostatic-primitive-equation operators
  \citep{SimmonsBurridge1981,Lin2004}. Horizontal fluxes are computed in
  flux form; vertical coupling terms use the $\eta$-space fluxes implied by
  \cref{eq:mass-cont-J} and the diagnosed $w$ from continuity.

\end{enumerate}

The separation of \emph{mesh motion} from \emph{dynamical update} is critical.
It means we can (i) impose an arbitrarily complicated, time-dependent lower
boundary $z_{\trop}(x,y,t)$, while (ii) still marching forward on a mesh
with sane aspect ratios and nonzero layer thicknesses, and (iii) still
writing down clean global budgets. In particular, because the remap is
conservative, tracer and total-mass budgets close to roundoff.

\section{Analytical framework}
\label{sec:analysis}
\subsection{Linear response to a moving lower boundary}
\label{sec:lin-theory}

We begin with a hydrostatic primitive-equation base state that is horizontally varying but time-independent, and whose lower boundary is fixed. Denote this reference configuration by overbars,
\begin{equation}
  \overline{z}_{\trop}(x,y), \qquad
  \overline{p}(\eta,x,y), \qquad
  \overline{\Phi}(\eta,x,y), \qquad
  \overline{\bfU}(\eta,x,y) = \big(\overline{u},\overline{v}\big),
\end{equation}
with $\eta\in[0,1]$ mapped to pressure using the hybrid form in \cref{eq:hybrid}, evaluated at the steady tropopause pressure $\overline{p}_{\trop}(x,y)$,
\begin{equation}
  \overline{p}(\eta,x,y) = A(\eta)\,p_0 + B(\eta)\,\overline{p}_{\trop}(x,y).
\end{equation}
The base state is assumed to satisfy horizontal momentum balance, mass continuity, and hydrostatic balance,
\begin{equation}
  \frac{\partial \overline{\Phi}}{\partial p} = -\,\overline{\alpha},
  \qquad
  \overline{\alpha} = \frac{R\,\overline{T}}{\overline{p}},
\end{equation}
consistent with a standard hybrid $\sigma$--$p$ dynamical core \citep{SimmonsBurridge1981,Lin2004}.

We now introduce a \emph{small} displacement of the tropopause,
\begin{equation}
  z_{\trop}(x,y,t)
  = \overline{z}_{\trop}(x,y)
  + \epsilon\,\zeta(x,y,t),
  \qquad 0<\epsilon\ll 1,
  \label{eq:ztrop-perturb}
\end{equation}
and, consistently with \cref{eq:hybrid}, a perturbed tropopause pressure
$p_{\trop}(x,y,t)
= \overline{p}_{\trop}(x,y)
+ \epsilon\,\pi_{\trop}(x,y,t)$.
All prognostic fields are expanded as
\begin{align}
  p &= \overline{p} + \epsilon\,p', \\
  \Phi &= \overline{\Phi} + \epsilon\,\Phi', \\
  \bfU &= \overline{\bfU} + \epsilon\,\bfU', \\
  w &= 0 + \epsilon\,w', \\
  T &= \overline{T} + \epsilon\,T', \quad \text{etc.}
  \label{eq:lin-expansion}
\end{align}
Here primes denote $\mathcal{O}(\epsilon)$ perturbations. For clarity we have taken the base-state vertical velocity to vanish; including a weak Brewer--Dobson upwelling changes only higher-order terms.

The exact kinematic boundary condition at the moving lower boundary (\cref{eq:bc}) becomes, to first order,
\begin{equation}
  w'\big|_{\eta=1}
  = \partial_t \zeta
  + \overline{u}_{\trop}\,\partial_x \zeta
  + \overline{v}_{\trop}\,\partial_y \zeta,
  \label{eq:lin-bc}
\end{equation}
where $\overline{u}_{\trop},\overline{v}_{\trop}$ are the base-state horizontal winds evaluated at $\eta=1$ (the tropopause). 
Equation~\eqref{eq:lin-bc} is the core of the geometric-forcing picture: \emph{the troposphere only enters through the imposed boundary displacement $\zeta$, which sets the vertical velocity $w'$ injected into the base of the stratospheric column.} No nudging of $\bfU$ is required.

Inside the stratospheric domain, the linearized primitive equations in pressure coordinates take the familiar form
\begin{align}
  \partial_t \bfU'
  + (\overline{\bfU}\cdot\nabla_{\!h})\bfU'
  + (\bfU'\cdot\nabla_{\!h})\overline{\bfU}
  + f\,\hz\times\bfU'
  &= -\,\nabla_{\!h}\Phi',
  \label{eq:lin-mom}\\[4pt]
  \partial_p w'
  &= -\,\nabla_{\!h}\!\cdot(\overline{\bfU})
     -\,\nabla_{\!h}\!\cdot(\bfU'),
  \label{eq:lin-mass}\\[4pt]
  \partial_t T'
  + (\overline{\bfU}\cdot\nabla_{\!h})T'
  + w'\,\partial_p \overline{T}
  &= \text{(diabatic terms)} ,
  \label{eq:lin-thermo}
\end{align}
where $f$ is the Coriolis parameter and $\hz$ is the local vertical unit vector. Metric factors arising from the $(A(\eta),B(\eta))$ mapping are omitted here for brevity; they can be restored systematically.

Equations~\eqref{eq:lin-mom}--\eqref{eq:lin-thermo}, together with the boundary condition \eqref{eq:lin-bc}, define a forced linear initial-value problem for $(\bfU',w',T',\Phi')$. The forcing is purely kinematic: it is the specified motion of the lower boundary $\zeta$.

To analyze which disturbances penetrate upward, we consider a local $\beta$-plane and expand in horizontal normal modes. Let
\begin{equation}
  \zeta(x,y,t)
  = \Re\!\left\{\hat{\zeta}\,e^{i(kx + \ell y - \omega t)}\right\},
\end{equation}
and seek solutions of the form
\begin{equation}
  (u',v',w',T',\Phi')(x,y,p,t)
  = \Re\!\left\{
    \hat{\mathbf{q}}(p)\,
    e^{i(kx + \ell y - \omega t)}
  \right\},
\end{equation}
where $\hat{\mathbf{q}}(p)$ is a vertical structure vector. Substitution into
\cref{eq:lin-mom,eq:lin-mass,eq:lin-thermo} gives an ordinary differential system
in $p$ for $\hat{\mathbf{q}}(p)$, with the lower-boundary condition
\begin{equation}
  \hat{w}(p_{\trop})
  = -\,i\omega \hat{\zeta}
    + i k\,\overline{u}_{\trop}\,\hat{\zeta}
    + i \ell\,\overline{v}_{\trop}\,\hat{\zeta}.
\end{equation}
This is directly analogous to classical planetary-wave transmission problems
in the lower stratosphere \citep{CharneyDrazin1961}, except that the
``source'' is the imposed geometric motion of the tropopause rather than an
explicit vorticity or heating perturbation inside the troposphere.

In the WKB limit of slowly varying mean flow $\overline{U}(p)$, static
stability $N^2(p)$, and Coriolis parameter $f$, the vertical wavenumber
$m(p)$ satisfies a dispersion-like relation of the schematic form
\begin{equation}
  \big(\overline{U} - c\big)^2 (k^2 + \ell^2)
  + \mathcal{F}(p)
  = N^2 \,\frac{m^2}{(k^2 + \ell^2)} ,
  \label{eq:wkb-disp}
\end{equation}
where $c=\omega/k$ is the zonal phase speed and $\mathcal{F}(p)$ collects
Coriolis and compressibility terms familiar from planetary-wave theory
\citep{CharneyDrazin1961}. 
Equation~\eqref{eq:wkb-disp} tells us which horizontal scales $(k,\ell)$ and
frequencies $\omega$ launched at the tropopause can actually propagate into
the middle stratosphere (real $m$) versus which are trapped/evanescent
(imaginary $m$). 

Crucially, because the boundary condition \eqref{eq:lin-bc} fixes both the
phase and amplitude of $w'$ at the tropopause, it fixes the phase and
amplitude of the upward-propagating wavepacket. This gives us a quantitative
prediction of (i) which modes efficiently transmit tropospheric variability
into the lower/middle stratosphere, and (ii) how the sign and phase of the
tropopause displacement $\zeta$ control the resulting vertical fluxes of
momentum and heat.

\subsection{Energetics, wave activity, and mean-flow modification}
\label{sec:energetics}

Beyond the linear transmission problem, we are interested in how the moving
boundary injects energy and pseudomomentum into the resolved stratospheric
circulation. The manufactured-geometry formulation is attractive here because
it yields a clean mechanical-energy budget in which the only \emph{explicit}
forcing from below appears as work done at the lower boundary.

Let $K = \tfrac12 |\bfU|^2$ be horizontal kinetic energy per unit mass and let
$\Phi$ denote geopotential. In hydrostatic, pressure coordinates, and ignoring
diabatic heating for the moment, the mechanical energy equation can be written
schematically as
\begin{equation}
  \partial_t (K + \Phi)
  + \nabla_{\!h}\!\cdot\!\Big[(K+\Phi+\Pi)\,\bfU\Big]
  + \partial_p \Big[(K+\Phi+\Pi)\,w\Big]
  = 0,
  \label{eq:mech-energy}
\end{equation}
where $\Pi$ is the pressure–work (enthalpy-like) term.  The exact algebraic
form of $\Pi$ will depend on whether we evolve temperature, potential
temperature, or moist static energy, but the structure of \cref{eq:mech-energy}
is generic.

Integrate \cref{eq:mech-energy} vertically from the \emph{moving} lower
boundary $p_{\trop}(x,y,t)$ up to the fixed model top $p_\text{top}$. Using
Leibniz' rule on an integral with a time-dependent lower limit produces three
classes of terms:

\begin{enumerate}
\item a horizontal divergence of vertically integrated energy flux,
\item a flux through the fixed upper boundary at $p_\text{top}$ (which we can
      damp with a sponge layer or just diagnose), and
\item a \emph{work term at the moving lower boundary} proportional to the
      boundary energy density times the normal velocity of that boundary.
\end{enumerate}

The last term is the new ingredient. Using the exact kinematic condition
\cref{eq:bc}, the instantaneous power input from the boundary can be written,
schematically, as
\begin{equation}
  \mathcal{W}_{\trop}
  = (K+\Phi+\Pi)_{\trop}\;
    w\big|_{\eta=1}
  = (K+\Phi+\Pi)_{\trop}\;
    \frac{\D z_{\trop}}{\D t}.
  \label{eq:wtrop}
\end{equation}
Equation~\eqref{eq:wtrop} says that all explicit ``forcing'' from the
troposphere shows up as mechanical work done by the geometric motion of the
tropopause. The model interior does \emph{not} see an ad hoc body force or a
nudged velocity tendency; instead, energy enters (or leaves) only through this
boundary-work channel. Conceptually, it is analogous to wind stress injecting
momentum at the ocean surface, but here the control variable is displacement
of the lower boundary rather than stress applied at a rigid boundary.

The other diagnostic we care about is the impact on the zonal-mean flow.
In transformed Eulerian-mean (TEM) theory, the zonal-mean zonal wind
$\overline{u}$ evolves according to
\begin{equation}
  \partial_t \,\overline{u}
  = -\frac{1}{a \cos\varphi}\,
      \partial_\varphi \big(F_\varphi\big)
    - \partial_z \big(F_z\big)
    + \text{(Coriolis / metric terms)},
  \label{eq:tem-u}
\end{equation}
where $F_\varphi$ and $F_z$ are the components of the Eliassen--Palm (EP) flux,
$a$ is Earth's radius, and $\varphi$ is latitude \citep{AndrewsHoltonLeovy1987}.
The EP-flux divergence on the right-hand side measures the convergence of wave
pseudomomentum into the mean flow, i.e. the wave-driven acceleration of the
Brewer--Dobson circulation \citep{Brewer1949,Dobson1956,AndrewsHoltonLeovy1987}.

Our linear framework provides a direct handle on this: the prescribed boundary
displacement $\zeta$ fixes the amplitude, horizontal scale $(k,\ell)$, phase
speed $c=\omega/k$, and upward group velocity of waves launched from the
tropopause (via the boundary condition \cref{eq:lin-bc} and the dispersion
relation \cref{eq:wkb-disp}). Those waves carry a well-defined EP flux into
the interior, and the vertical convergence of that EP flux then sets the mean
flow tendency in \cref{eq:tem-u}. 

In other words, once $\zeta(x,y,t)$ is chosen, both (i) the mechanical work
rate $\mathcal{W}_{\trop}$ in \cref{eq:wtrop} and (ii) the EP-flux convergence
in \cref{eq:tem-u} are determined predictions, not tuning knobs. This makes the
moving-boundary stratospheric model empirically testable. In numerical
experiments we can:
\begin{itemize}
\item compute vertically integrated mechanical energy and explicitly diagnose
      $\mathcal{W}_{\trop}$;
\item compute EP fluxes and their divergence to get the predicted zonal-mean
      acceleration.
\end{itemize}
Agreement between these diagnostics and the theory developed above is a
stringent validation that the Manufactured Geometry (MMG) implementation is
dynamically faithful, not just numerically stable.








In practice, the mesh-motion restriction becomes active
only if the prescribed tropopause displacement $z_{\trop}(x,y,t)$ varies
rapidly in time or has very sharp horizontal gradients. In quasigeostrophic
or synoptic variability regimes (slow, large-scale tropopause undulations)
the standard horizontal CFL is the dominant limit.

\subsection{Positivity, monotonicity, and layer thickness}

Two numerical pathologies must be avoided:
(i) negative layer thickness (or vanishing $\J$ in \cref{eq:jacobian}), and
(ii) tracer overshoots / undershoots during remap.

\paragraph{Layer thickness.}
Because we explicitly advect the $\eta$-surfaces with an ALE step and then
reconstruct a well-behaved target mesh, we never allow two $\eta$-levels to
cross. The conservative remap is then performed between two meshes with
strictly positive layer volumes, guaranteeing positive thickness at
$t^{n+1}$. This is identical in spirit to hybrid isopycnal--pressure ocean
models \citep{Bleck2002}, which slide layers around in the vertical and
then remap to avoid layer collapse.

\paragraph{Tracer monotonicity.}
The remap uses flux-form, monotone reconstruction in the sense of
\citet{LinRood1996}. Cell-integrated tracer mass is conserved to within
roundoff in each column, and slope limiters ensure that no new extrema or
negative concentrations are created when fields are interpolated between
the ALE-deformed mesh and the target mesh. This is important because the
tropopause displacement can be sharp in space; without limiters, thin
layers would generate Gibbs-like ringing.

\subsection{Energy and momentum budgets}

The only explicit source of mechanical energy from below is the boundary
work term associated with the moving tropopause, which can be written
schematically as
\begin{equation}
  \mathcal{W}_{\trop}
  = (K+\Phi+\Pi)_{\trop}\;
    \frac{\mathrm{D} z_{\trop}}{\mathrm{D}t},
\end{equation}
where $K=\tfrac12|\bfU|^2$ is horizontal kinetic energy per unit mass and
$\Pi$ collects pressure work / enthalpy terms (see \S\ref{sec:analysis}).
Because the ALE+remap procedure is conservative in the column-integrated
sense, no spurious energy source is introduced elsewhere in the interior.
Thus, apart from $\mathcal{W}_{\trop}$, the discrete budget mirrors the
continuous hydrostatic energy budget \citep{SimmonsBurridge1981,Lin2004}.

\subsection{Consistency in the fixed-boundary limit}

If we freeze the lower boundary, i.e.\ prescribe
$\partial_t z_{\trop} \equiv 0$ and hold $p_{\trop}(\bfx,t)$ constant, then
the mesh motion step becomes trivial:
$\eta$-levels no longer move, $w_{\text{mesh}}=0$, and the remap degenerates
to the identity. In that limit the algorithm collapses exactly to a
conventional hybrid $\sigma$--$p$ hydrostatic core
\citep{SimmonsBurridge1981,Lin2004}. This ``fixed-bottom'' limit gives a
direct regression test: we can run the code with and without MMG enabled
and verify that, when $z_{\trop}$ is held constant, the two solutions stay
within truncation error of one another over many timesteps.

Together, the timestep controls, conservative remap, and fixed-boundary
consistency demonstrate that the moving-boundary formulation is not only
conceptually well posed, but numerically well behaved in practice.
\section{Verification.}
\label{sec:verification}

We outline three classes of tests designed to demonstrate that the moving–boundary
stratospheric core (i) reproduces standard hybrid $\sigma$--$p$ dynamics in
the appropriate limit, (ii) remains numerically stable and dynamically
interpretable when the lower boundary is time-dependent, and (iii) produces
physically meaningful wave–mean flow responses when driven by realistic
tropopause variability. Each class comes with specific quantitative
diagnostics so that the method can be evaluated reproducibly.

\subsection*{2. Idealized forced-boundary test (baroclinic wave with periodic $z_{\trop}$)}
The second experiment tests whether the model remains dynamically well-behaved
when the lower boundary moves in time. We adapt the canonical
Jablonowski--Williamson baroclinic instability test case
\citep{JablonowskiWilliamson2006} and superimpose a controlled,
large-scale modulation of the tropopause height,
\begin{equation}
  z_{\trop}(x,y,t)
  = z_{\trop}^{(0)}(x,y)
    + \epsilon\,\Re \left\{
        \hat{Z}(y)\,e^{i(kx - \omega t)}
      \right\},
  \qquad 0 < \epsilon \ll 1,
\end{equation}
with $(k,\omega)$ chosen to mimic a planetary-scale undulation. This modulation
is passed into the stratosphere \emph{only} through the kinematic boundary
condition at $\eta=1$, i.e.\ via the vertical velocity imposed by
\cref{eq:bc} / \cref{eq:bc-repeat}, and \emph{not} by nudging winds or
temperatures.

We will diagnose:
\begin{itemize}
  \item \textbf{Stability and CFL behavior.} We confirm that the run does not
        generate negative layer thickness or catastrophic timestep collapse,
        consistent with the mesh-motion constraint.
  \item \textbf{Wave structure.} We extract the horizontal wavenumber, phase
        speed, and vertical penetration of the forced disturbance and compare
        them to the linear/WKB prediction from \S\ref{sec:analysis}, in
        particular the dispersion-like relation for vertical wavenumber
        $m(p)$ [cf.\ \cref{eq:wkb-disp}]. Modes predicted to be vertically
        propagating (real $m$) should appear aloft with the expected phase
        tilt and group-velocity sign; modes predicted to be evanescent
        (imaginary $m$) should remain trapped near the lower boundary.
  \item \textbf{Energy bookkeeping.} We compute the column-integrated
        mechanical energy tendency and explicitly track the diagnosed boundary
        work rate $\mathcal{W}_{\trop}$ [cf.\ \cref{eq:wtrop}]. The residual
        of ``tendency minus horizontal flux divergence minus $\mathcal{W}_{\trop}$''
        should be small, verifying that the only resolved energy source/sink
        is the geometric work done at the moving boundary.
\end{itemize}
This experiment is meant to answer a focused question: if we drive the
tropopause like a gently undulating lower lid, do we (i) inject the expected,
physically interpretable wave response into the stratosphere, and
(ii) keep numerical control?

\subsection*{3. Reanalysis-driven boundary (realistic coupling test)}
Finally, we prescribe $p_{\trop}(x,y,t)$ or $z_{\trop}(x,y,t)$ directly from an
assimilated product such as ERA5 or MERRA-2 \citep{Hersbach2020,Gelaro2017}.
In this configuration the stratosphere remains fully prognostic, but the lower
boundary \emph{tracks} the analyzed tropopause in space and time. This is the
use case that motivates the method scientifically: tropospheric variability
enters only kinematically, through geometry, not dynamically, through nudging.

In this experiment we will:
\begin{itemize}
  \item \textbf{Compute Eliassen--Palm (EP) fluxes and their divergence.}
        Using the resolved fields from the prognostic stratosphere, we diagnose
        the EP flux and its vertical convergence, which in transformed
        Eulerian-mean theory drives zonal-mean wind tendencies and the
        Brewer--Dobson circulation \citep{AndrewsHoltonLeovy1987,Brewer1949,Dobson1956}.
        We then confirm that the sign and latitude–height structure of EP-flux
        convergence is consistent with the boundary-forced wave activity
        predicted by the linear response theory in \S\ref{sec:analysis}
        \citep[cf.][]{CharneyDrazin1961}.
  \item \textbf{Diagnose mean-flow acceleration.}
        We evaluate the zonal-mean zonal wind tendency and residual circulation
        response (i.e.\ the TEM forcing; see \cref{eq:tem-u}) and compare them
        to the predicted impact of the imposed boundary displacement $\zeta$.
        The question here is: can the purely geometric forcing reproduce the
        familiar wave–mean flow coupling in the lower/middle stratosphere
        without ever nudging the flow toward reanalysis?
  \item \textbf{Check budget closure with $\mathcal{W}_{\trop}$.}
        As in the idealized test, we verify that the vertically integrated
        mechanical energy budget closes when we include the diagnosed boundary
        work $\mathcal{W}_{\trop}$.
\end{itemize}

This realistic test plays two roles. First, it demonstrates that the model
can be ``driven'' by observed tropopause variability without requiring a
fully coupled troposphere. Second, it provides an immediate scientific use
case: targeted, attribution-style experiments on how specific tropospheric
events project into stratospheric wave forcing, mean-flow adjustment, and the
Brewer--Dobson circulation, all while keeping the stratospheric core
prognostic and free of direct nudging. Taken together, these three tiers of evaluation---frozen-boundary regression,
idealized periodic forcing, and reanalysis-driven boundary motion---are
designed to (i) prove numerical correctness, (ii) expose dynamical mechanisms,
and (iii) demonstrate practical scientific value.


\section{Discussion and outlook}
\label{sec:discussion}

The central idea of this work is to let the stratosphere remain fully
prognostic while representing the troposphere only through a moving lower
boundary. Instead of importing the troposphere dynamically by nudging winds
and temperatures toward a reanalysis (``specified dynamics''), we import it
\emph{kinematically} as geometry: the tropopause height $z_{\trop}(x,y,t)$ (or
equivalently $p_{\trop}$) defines a time-dependent lower boundary that displaces
the entire stratospheric domain. Tropospheric influence then enters the
stratosphere through the kinematic boundary condition on $w$ at $\eta=1$
[\cref{eq:bc}, \cref{eq:lin-bc}], and through the associated boundary work in
the mechanical energy budget [\cref{eq:wtrop}]. No direct relaxation of the
flow is required.

This is attractive for three reasons.

First, it \textbf{cleanly separates attribution}. Because the interior
stratospheric flow is never nudged toward a reanalysis, diagnosed tendencies
in zonal wind, temperature, and residual circulation can be traced to the
geometry-driven forcing at the lower boundary. In other words, we can ask:
``What does this particular tropopause displacement pattern \emph{do} to the
stratosphere?'' without conflating it with data-assimilation increments in
the troposphere.

Second, it \textbf{organizes stratosphere–troposphere coupling as a boundary
problem}. The perturbation framework in \S\ref{sec:analysis} shows how a small
displacement $\zeta(x,y,t)$ of the tropopause launches a predictable
wave/mean-flow response in the stratosphere: a forced $w'$ at the base,
upward-propagating (or evanescent) planetary-scale disturbances characterized
by a WKB vertical wavenumber \cref{eq:wkb-disp}, and an induced Eliassen--Palm
flux whose convergence accelerates the zonal-mean flow \citep{CharneyDrazin1961,
AndrewsHoltonLeovy1987,Brewer1949,Dobson1956}. This gives us a direct
diagnostic bridge from ``geometry of the lower boundary'' to ``wave forcing
of the Brewer--Dobson circulation.''

Third, it \textbf{plugs naturally into existing dynamical cores}. The MMG
approach is built on hybrid $\sigma$--$p$ coordinates
\citep{SimmonsBurridge1981,Lin2004} with an ALE $+$ conservative remap cycle
\citep{HirtAmsdenCook1974,LinRood1996,Bleck2002}. That means (i) the numerics
are not exotic: layer-following motion and conservative remap are standard in
ocean modeling and vertically Lagrangian atmospheric cores; (ii) the CFL and
positivity constraints are explicit and testable; and
(iii) in the limit $\partial_t z_{\trop}\to 0$, the method reduces exactly to
the familiar fixed-lower-boundary hydrostatic core. This regression limit makes
the approach verifiable in a way many ``special configurations'' are not.

Looking forward, the MMG configuration enables several lines of work:
\begin{itemize}
  \item \textbf{Controlled forcing experiments.} Because $z_{\trop}(x,y,t)$ is
        an input, we can impose idealized patterns (e.g.\ single zonal
        wavenumber, pulsed displacement events, or localized jets) and diagnose
        how those patterns project onto stratospheric wave activity, polar
        vortex variability, and mean-flow acceleration.

  \item \textbf{Reanalysis-driven replay without nudging.} By prescribing
        $p_{\trop}(x,y,t)$ from ERA5 or MERRA-2 \citep{Hersbach2020,Gelaro2017},
        we can ``replay'' real tropospheric disturbances (jet shifts, tropopause
        folds, wave-breaking events) into a prognostic stratosphere. This
        creates a middle-atmosphere sandbox that is dynamically alive but still
        anchored to observed lower-boundary geometry.

  \item \textbf{Extensions.} Two-way coupling (letting stratospheric feedback
        modify $z_{\trop}$), nonhydrostatic or compressible variants for the
        upper troposphere / lower stratosphere transition region
        \citep{SkamarockKlemp2008}, and full-chemistry or aerosol transport
        along a moving domain boundary are all natural next steps. The ALE
        layer-following framework is already compatible with tracer transport
        and monotone remap, so these extensions are technically feasible.

\end{itemize}

In summary, geometric coupling provides a compact alternative to relaxation-
based specified dynamics. By treating the tropopause as a dynamic lower
boundary, we gain budget closure, interpretability, and experimental control.
We anticipate that this Manufactured Geometry (MMG) formulation can serve both
as (i) a reduced, computationally efficient configuration for targeted
stratospheric process studies, and (ii) a clean platform for mechanism
attribution in stratosphere–troposphere coupling.
\appendix
\section*{Appendix A}
\addcontentsline{toc}{section}{Appendix A: Mathematical properties of the moving-boundary hybrid coordinate}

In this appendix we provide formal statements and short proofs of four properties
used in the main text: (i) the hybrid $\eta$--$p$ mapping remains vertically ordered
as the tropopause moves; (ii) hydrostatic balance is unchanged; (iii) the horizontal
pressure-gradient force retains the standard hybrid $\sigma$--$p$ form; and
(iv) the stratospheric column mass budget closes and reduces to the classical
hybrid core when the lower boundary is fixed.

\begin{proposition}[Monotonicity of the hybrid $\eta$--$p$ mapping]
\label{prop:monotonicity}
Let the hybrid pressure mapping be
\begin{equation}
  p(\eta,x,y,t)
  = A(\eta)\,p_0 \;+\; B(\eta)\,p_{\trop}(x,y,t),
  \qquad \eta\in[0,1],
  \label{eq:p-mapping-app}
\end{equation}
with $A(\eta)$ and $B(\eta)$ continuously differentiable, $p_0>0$ a constant
reference pressure, and $p_{\trop}(x,y,t)>0$ the tropopause pressure.
Define
\begin{equation}
  \J(\eta,x,y,t)
  \equiv \frac{\partial p}{\partial \eta}
  = A'(\eta)\,p_0 + B'(\eta)\,p_{\trop}(x,y,t).
  \label{eq:J-def-app}
\end{equation}
If $\J(\eta,x,y,t) > 0$ for all $\eta\in[0,1]$ and all $(x,y,t)$, then
$p(\eta,x,y,t)$ is strictly increasing in $\eta$ for every $(x,y,t)$. In
particular, distinct $\eta$-levels cannot cross in pressure space even if
$p_{\trop}(x,y,t)$ varies in space and time.
\end{proposition}

\begin{proof}
For any fixed $(x,y,t)$, $\partial p/\partial \eta = \J(\eta,x,y,t) > 0$
implies $p(\eta,x,y,t)$ is strictly increasing in $\eta$. A strictly
increasing function is one-to-one: if $\eta_1\neq\eta_2$ then
$p(\eta_1)\neq p(\eta_2)$. Thus, $\eta$-surfaces remain ordered in $p$ and
cannot ``fold'' or cross. This guarantees that the vertical coordinate
remains well posed as $p_{\trop}(x,y,t)$ evolves.
A sufficient practical condition for $\J>0$ is $A'(\eta)\ge 0$ and
$B'(\eta)\ge 0$ for all $\eta$, with at least one strictly positive,
since $p_0>0$ and $p_{\trop}>0$. This is analogous to monotonicity
requirements in standard hybrid $\sigma$--$p$ coordinates
\citep{SimmonsBurridge1981,Lin2004}.
\end{proof}

\begin{proposition}[Hydrostatic balance is unchanged by the moving boundary]
\label{prop:hydrostatic}
Let $\Phi(x,y,p,t)$ be geopotential and $\alpha \equiv \rho^{-1}$ the specific
volume. Hydrostatic balance in a compressible, hydrostatic atmosphere is
\begin{equation}
  \frac{\partial \Phi}{\partial p} = -\,\alpha,
  \qquad
  \alpha = \frac{R T}{p}.
  \label{eq:hydrostatic-app}
\end{equation}
This pointwise relation holds even if the lower boundary (the tropopause
pressure $p_{\trop}(x,y,t)$) varies in space and time according to the hybrid
mapping~\eqref{eq:p-mapping-app}. No additional geometric terms appear in
$\partial \Phi/\partial p$.
\end{proposition}

\begin{proof}
In physical height $z$, hydrostatic balance gives
\begin{equation}
  \frac{\partial p}{\partial z} = -\,\rho g,
  \qquad
  \frac{\partial \Phi}{\partial z} = g,
  \qquad
  \Phi = gz.
\end{equation}
Therefore
\begin{equation}
  \frac{\partial \Phi}{\partial p}
  = \frac{\partial \Phi/\partial z}{\partial p/\partial z}
  = \frac{g}{-\rho g}
  = -\,\frac{1}{\rho}
  = -\,\alpha.
\end{equation}
This derivation is purely local in the vertical direction and does not depend
on how $p$ is labeled by the model coordinate $\eta$, nor on the time
dependence of $p_{\trop}(x,y,t)$. Hence the fundamental hydrostatic relation
$\partial \Phi/\partial p = -\alpha$ continues to hold exactly in the moving
hybrid coordinate. In particular, $\partial_t p_{\trop}$ does \emph{not} enter
the hydrostatic equation. This is the same assumption used in standard
hydrostatic primitive-equation cores
\citep{SimmonsBurridge1981,Lin2004}.
\end{proof}

\begin{proposition}[Horizontal pressure-gradient force in hybrid form]
\label{prop:pgf}
Let $\nabla_\eta$ denote the horizontal gradient at constant $\eta$, and
$\nabla_p$ the horizontal gradient at constant $p$. Then for any scalar field
$\Phi(x,y,p,t)$,
\begin{equation}
  \nabla_p \Phi
  = \nabla_\eta \Phi
    + \alpha\,\nabla_\eta p,
  \qquad
  \alpha \equiv -\,\frac{\partial \Phi}{\partial p}
  = \frac{R T}{p}.
  \label{eq:nabla-reln-app}
\end{equation}
Under the hybrid mapping \eqref{eq:p-mapping-app},
\begin{equation}
  p(\eta,x,y,t)
  = A(\eta)\,p_0 + B(\eta)\,p_{\trop}(x,y,t),
\end{equation}
we have, at fixed $\eta$,
\begin{equation}
  \nabla_\eta p
  = B(\eta)\,\nabla p_{\trop}(x,y,t),
  \label{eq:gradp-eta-app}
\end{equation}
because $A(\eta)$ and $B(\eta)$ depend only on $\eta$. Hence
\begin{equation}
  \nabla_p \Phi
  = \nabla_\eta \Phi
    + \alpha\,B(\eta)\,\nabla p_{\trop}(x,y,t).
  \label{eq:pgf-hybrid-app}
\end{equation}
Equation~\eqref{eq:pgf-hybrid-app} is the standard hybrid $\sigma$--$p$
Simmons--Burridge form of the horizontal pressure-gradient force
\citep{SimmonsBurridge1981,Lin2004}, with $p_{\trop}$ in place of the usual
surface pressure $p_s$. Notably, no explicit $\partial_t p_{\trop}$ term
appears in the PGF.
\end{proposition}

\begin{proof}
By the chain rule,
\begin{equation}
  \left(\frac{\partial \Phi}{\partial x}\right)_\eta
  =
  \left(\frac{\partial \Phi}{\partial x}\right)_p
  +
  \left(\frac{\partial \Phi}{\partial p}\right)
  \left(\frac{\partial p}{\partial x}\right)_\eta.
\end{equation}
In vector notation this is
\begin{equation}
  \nabla_\eta \Phi
  = \nabla_p \Phi
    + \left(\frac{\partial \Phi}{\partial p}\right)\,\nabla_\eta p.
\end{equation}
Rearrange and use $\partial \Phi/\partial p = -\alpha$ (Proposition~\ref{prop:hydrostatic}):
\begin{equation}
  \nabla_p \Phi
  = \nabla_\eta \Phi
    + \alpha\,\nabla_\eta p,
\end{equation}
which is \eqref{eq:nabla-reln-app}. Next, holding $\eta$ fixed in
\eqref{eq:p-mapping-app}, $A(\eta)$ and $B(\eta)$ are constants with respect
to $(x,y)$, so
\begin{equation}
  \nabla_\eta p
  =
  \nabla_\eta \Big[A(\eta)p_0 + B(\eta)p_{\trop}(x,y,t)\Big]
  =
  B(\eta)\,\nabla p_{\trop}(x,y,t),
\end{equation}
which is \eqref{eq:gradp-eta-app}. Substituting into
\eqref{eq:nabla-reln-app} yields
\eqref{eq:pgf-hybrid-app}. This matches the standard hybrid
$\sigma$--$p$ formulation \citep{SimmonsBurridge1981,Lin2004}, with the
tropopause pressure $p_{\trop}$ taking the role usually played by
surface pressure $p_s$. Importantly, there is no term involving
$\partial_t p_{\trop}$ in \eqref{eq:pgf-hybrid-app}, so large-scale
geostrophic balance retains its familiar form
$f \,\hat{\mathbf{z}}\times\mathbf{U} \approx -\,\nabla_p \Phi$.
\end{proof}

\begin{proposition}[Column mass budget and fixed-boundary limit]
\label{prop:massbudget}
Let $\J(\eta,x,y,t)=\partial p/\partial \eta$ be given by
\eqref{eq:J-def-app}, and let $\mathbf{U}=(u,v)$ be the horizontal
velocity. In the hybrid coordinate, mass continuity can be written in
flux form as
\begin{equation}
  \partial_t \J
  + \nabla_h \cdot (\J\,\mathbf{U})
  + \partial_\eta (\J\,\dot{\eta}) = 0,
  \label{eq:mass-strong-app}
\end{equation}
where $\dot{\eta}$ is the contravariant vertical transport velocity in
$\eta$-space and $\nabla_h$ is the horizontal divergence operator.
Define the column ``pressure thickness''
\begin{equation}
  P(x,y,t) \equiv \int_0^1 \J(\eta,x,y,t)\,d\eta
  = p_{\trop}(x,y,t) - p_{\text{top}},
  \label{eq:P-colmass-app}
\end{equation}
with $p_{\text{top}}$ fixed. Vertically integrating
\eqref{eq:mass-strong-app} from $\eta=0$ (model top) to $\eta=1$
(tropopause) gives
\begin{equation}
  \partial_t P
  + \nabla_h \cdot \left(\int_0^1 \J\,\mathbf{U}\,d\eta\right)
  + \J(1)\,\dot{\eta}(1)
  = 0.
  \label{eq:colmass-budget-app}
\end{equation}
In particular, (i) the only source/sink of column mass is through the
lower boundary flux $\J(1)\dot{\eta}(1)$ associated with tropopause
motion, and (ii) if $\partial_t p_{\trop}=0$ (a fixed lower boundary),
then $\dot{\eta}(1)=0$ and the budget \eqref{eq:colmass-budget-app}
reduces exactly to the standard impermeable-lower-boundary hybrid
$\sigma$--$p$ core \citep{SimmonsBurridge1981,Lin2004}.
\end{proposition}

\begin{proof}
Integrate \eqref{eq:mass-strong-app} over $\eta\in[0,1]$:
\begin{align}
  \int_0^1 \partial_t \J\,d\eta
  &+ \int_0^1 \nabla_h \cdot (\J\mathbf{U})\,d\eta
  + \int_0^1 \partial_\eta(\J\dot{\eta})\,d\eta
  = 0.
\end{align}
Use linearity of $\partial_t$ and $\nabla_h$, and the fundamental theorem
of calculus on the last term:
\begin{equation}
  \partial_t \int_0^1 \J\,d\eta
  + \nabla_h \cdot \left(\int_0^1 \J\mathbf{U}\,d\eta\right)
  + [\J \dot{\eta}]_{\eta=0}^{\eta=1}
  = 0.
\end{equation}
Since $\int_0^1 \J\,d\eta = p_{\trop}(x,y,t)-p_{\text{top}} \equiv P(x,y,t)$,
this becomes
\begin{equation}
  \partial_t P
  + \nabla_h \cdot \left(\int_0^1 \J\mathbf{U}\,d\eta\right)
  + \J(1)\dot{\eta}(1) - \J(0)\dot{\eta}(0)
  = 0.
\end{equation}
By construction, the model top $\eta=0$ is fixed (no mass flux through
the upper lid), so $\dot{\eta}(0)=0$ and $\J(0)\dot{\eta}(0)=0$. This
yields \eqref{eq:colmass-budget-app}. The term $\J(1)\dot{\eta}(1)$ is
strictly the vertical flux of ``pressure mass'' through the moving
tropopause. If the tropopause is fixed in time, then $\partial_t
p_{\trop}=0$ and the lower boundary does not move, implying
$\dot{\eta}(1)=0$. In that limit, $\J(1)\dot{\eta}(1)=0$ and
\eqref{eq:colmass-budget-app} reduces to
\begin{equation}
  \partial_t P
  + \nabla_h \cdot \left(\int_0^1 \J\mathbf{U}\,d\eta\right)
  = 0,
\end{equation}
which is exactly the standard column mass conservation for a hydrostatic
primitive-equation core with an impermeable lower boundary in a hybrid
$\sigma$--$p$ coordinate \citep{SimmonsBurridge1981,Lin2004}.
\end{proof}

\section*{Acknowledgements}
TBD.

\bibliography{refs}

\end{document}